\documentstyle[preprint,eqsecnum,aps]{revtex}

\begin{document}
\draft
\preprint{}
\title{Magnon Damping by magnon-phonon coupling in Manganese
Perovskites\\}
\author{Pengcheng Dai,$^1$ H.Y. Hwang,$^2$ Jiandi
Zhang,$^{1,}$\cite{byline} J. A.
Fernandez-Baca,$^1$ S-W. Cheong,$^{2,3}$ C. Kloc,$^2$ Y. Tomioka,$^4$ and
Y. Tokura$^{4,5}$}
\address{$^1$Oak Ridge National Laboratory, Oak Ridge, Tennessee 37831-6393}
\address{$^2$Bell Laboratories, Lucent Technologies, Murray Hill, New
Jersey 07974}
\address{$^3$Department of Physics and Astronomy, Rutgers University,
Piscataway, New Jersey 08855}
\address{$^4$Joint Research Center for Atom Technology (JRCAT), Tsukuba
305-8562, Japan}
\address{$^5$Department of Applied Physics, University of Tokyo, Tokyo
113-8656, Japan}
\date{\today}
\maketitle
\begin{abstract}
Inelastic neutron scattering was used to systematically investigate the
spin-wave
excitations (magnons) in ferromagnetic manganese perovskites. In spite of the
large differences in the
Curie temperatures ($T_C$s) of different manganites,
their low-temperature
spin waves were found to have very
similar dispersions with the zone boundary magnon softening.
From the wavevector dependence of the magnon lifetime effects and its
correlation with
the dispersions of the optical phonon modes, we argue that a strong
magneto-elastic
coupling is responsible for the observed low temperature
anomalous spin dynamical behavior
of the manganites.
\end{abstract}


\narrowtext
The elementary magnetic excitations (spin waves)
in a ferromagnet can provide direct
information about the itinerancy of the unpaired electrons contributing to the
ordered moment.  In insulating (local moment) ferromagnets, such
excitations
are usually well defined throughout the Brillouin zone and can be described
by the
Heisenberg model of magnetism \cite{heisenberg}. On the other hand,
metallic (itinerant)
ferromagnets are generally characterized by the disappearance of spin waves
at finite
energy and momentum due to the presence of the Stoner (electron-hole
pair) excitation continuum associated with the band
structure and itinerant electrons in the system \cite{izuyama}.  In the
ferromagnetic
 manganese perovskites (manganites)
$A_{0.7}B_{0.3}$MnO$_{3}$ (where $A$ and $B$ are rare-earth and
alkaline-earth elements respectively.) \cite{jonker},
the ferromagnetism and electric conductivity
can be continually suppressed by different $A$($B$) substitutions
until an insulating, charge-ordered ground
state is
stabilized \cite{hwang1}.
Approaching this insulating ground state (with decreasing  Curie
temperature $T_C$), the
low  temperature spin-wave
dispersions  have been found to
deviate \cite{hwang2} from the nearest-neighbor Heisenberg exchange
Hamiltonian that has been successfully
applied to the higher $T_C$ materials \cite{martin,toby,endoh,moudden}.
However, the microscopic origin
for such deviation is unknown.
Furthermore, spin waves for the lower $T_C$ $A_{0.7}B_{0.3}$MnO$_{3}$
have very similar spin-wave
stiffness constants ($D$), contrary to the expectation of a
Heisenberg ferromagnet where $D$ is related to $T_C$ \cite{jeff,fernandez}.
Here
we argue that the deviation from the simple Heisenberg Hamiltonian and the
observation of magnon damping at large wavevectors are due to
strong magneto-elastic (or magnon-phonon) interactions,
consistent with the electron-lattice coupling \cite{zhou}
for the manganites
in the proximity of the charge ordered insulating state.

We study
$A_{0.7}B_{0.3}$MnO$_{3}$ manganites because these  materials exhibit a large
resistivity drop that is intimately related to the
paramagnetic-to-ferromagnetic phase
transition at $T_C$ \cite{jonker}.
Due to the octahedral
crystalline field, the 3$d$ energy level of the Mn ion in
$A_{0.7}B_{0.3}$MnO$_{3}$ is split into a
low-lying $t_{2g}$ triplet and
a higher energy $e_g$ doublet.  In 1951, Zener \cite{zener} postulated that
the conductivity in these mixed valence systems
  was due to the simultaneous hopping of an ($e_g$) electron (with
electron transfer energy $t$)
 from Mn$^{3+}$ ($3d^4$) to the connecting O$^{2-}$ and from the O$^{2-}$
to the Mn$^{4+}$ ($3d^3$) $e_g$ band; hence termed double-exchange (DE).
Because all the electrons in the Mn 3$d$ levels are polarized by a strong
intra-atomic exchange interaction $J_H$ (Hund-rule coupling), such hopping
tends to align the spins of Mn ions on adjacent sites parallel
\cite{anderson}.
 Thus, the DE mechanism
provided a qualitative interpretation of coupled ferromagnetic ordering and
 electric conductivity \cite{jonker}.
However, recent theoretical work \cite{millis,bishop} suggests that DE
mechanism alone
cannot explain the temperature dependence of the resistivity near and above
$T_C$.
Additional interactions, such as a strong
dynamical Jahn-Teller (JT) based
electron-lattice coupling, are necessary
 to explain the magnitude of the resistivity drop across $T_C$.
 In this scenario, the JT distortion of the Mn$^{3+}$O$_6$ octahedra lifts
the double degeneracy of
 $e_g$ electrons in Mn$^{3+}$ and
causes a local lattice distortion that may trap carriers to form
a polaron. The
formation of lattice polarons in the paramagnetic state leads to the
localization
of conduction band ($e_g$) electrons above $T_C$ and
hence the insulating behavior. On cooling below $T_C$ however, the growing
ferromagnetic order
increases the $e_g$ electron hopping (kinetic) energy $t$ and decreases the
electron-lattice coupling strength sufficiently that metallic behavior occurs.
The system crosses over from a polaron regime to a
Fermi-liquid regime and the DE mechanism ultimately prevails in the low
temperature
metallic state \cite{millis}.

The electron-lattice coupling described above is dynamical, $i.e.$, it involves
vibrational distortions of the oxygen octahedron around the
Mn$^{3+}$ site. However,
the static lattice distortion, present because of the ionic size differences
of $A$ and $B$
in $A_{0.7}B_{0.3}$MnO$_{3}$, may also affect the electron hopping and the
Curie temperature $T_C$.
Therefore,  systematic studies of the spin-wave excitations  in
$A_{0.7}B_{0.3}$MnO$_{3}$ with decreasing $T_C$ and increasing
residual resistivity \cite{hwang1}
should reveal how the system evolves from an itinerant to a localized
ferromagnet. Such investigation
will also test to what extent the low temperature magnetic properties
of $A_{0.7}B_{0.3}$MnO$_{3}$
ferromagnets can be explained by the DE
mechanism.
For these purpose,
we have characterized the low temperature spin dynamics of the ferromagnetic
single crystals of Pr$_{0.63}$Sr$_{0.37}$MnO$_3$
(PSMO),
La$_{0.7}$Ca$_{0.3}$MnO$_3$ (LCMO), and Nd$_{0.7}$Sr$_{0.3}$MnO$_3$ (NSMO)
which have approximately the same nominal
carrier concentration but significantly different $T_C$s and
residual conductivity (see Figure 1).

The temperature dependent resistivity $\rho(T)$ for these three samples is
shown
in Figure 1.  The characteristic drop in $\rho$ coincident with $T_C$ is
clearly
seen to increase with decreasing $T_C$. At the same time
the residual resistivity increases almost linearly
with decreasing $T_C$, indicating that with increasing magnetoresistance
effect
the system becomes a worse metal at low temperatures.
An interesting feature of $\rho(T)$ at low temperatures is that all three
samples exhibit the same temperature dependence below $\sim$100 K
when $\rho(T)$ is scaled to the residual value $\rho(0)$.
The inset to Figure 1 illustrates this point.

The open symbols in Figure 2 summarize the spin-wave
dispersions along the $[0,0,\xi]$,
$[\xi,\xi,0]$, and $[\xi,\xi,\xi]$ directions
at 10 K for PSMO, LCMO, and NSMO \cite{note1}.  Clearly, the
dispersions of
these three manganites are remarkably similar at the measured frequencies,
suggesting that the magnetic exchange coupling strength, derived
from the hopping of the $e_g$ electrons between the Mn$^{3+}$ and Mn$^{4+}$
sites,
depend only weakly on $T_C$.
These results are in sharp contrast to the single-band
DE model where the spin-wave dispersions
are directly related to electronic bandwidth and hence
$T_C$ \cite{furukawa,okabe,kaplan,wang}.
In the strong coupling limit ($J_H/t\rightarrow\infty$) of this model,
the spin-wave dispersion of
the ferromagnet is consistent with the nearest-neighbor
Heisenberg Hamiltonian
and $D$ should be proportional to the electron transfer energy
$t$.
Previous work has shown that such single-band DE model is
adequate for describing the spin dynamics of the highest $T_C$
ferromagnetic manganites
\cite{martin,toby,endoh,moudden}. To estimate the spin-spin exchange coupling
strength, we note that the low-frequency spin waves of
 $A_{0.7}B_{0.3}$MnO$_3$ manganites LCMO \cite{jeff}, NSMO, and
PSMO
\cite{fernandez} are isotropic and gapless with a stiffness
$D\approx 165$ meV\AA $^{-2}$. For a simple cubic Heisenberg ferromagnet
with nearest-neighbor
exchange coupling $J$, $D=2JSa^2$, where $S$ is the magnitude of the electronic
spin at the magnetic ionic sites and $a$ is the lattice parameter.
From the measured spin-wave stiffness, one can calculate the exchange
coupling strength
 $J$ and hence the expected dispersion for a simple nearest-neighbor
Heisenberg ferromagnet.
The solid lines in Figure 2 show the outcome of such calculation which
 clearly misses the measured
spin-wave energies at large wavevectors.

Figure 3 shows typical constant-${\bf q}$ scans
along the $[0,0,\xi]$ and $[\xi,\xi,0]$
directions for LCMO and NSMO. Most of the data are well described by
Gaussian fits
which give the
amplitude, widths, and peak positions of the excitations. While the
dispersion curves shown in
Figure 2 are obtained by peak positions at different wavevectors, the
amplitude and widths
provide information about the damping and lifetime of the magnon
excitations.
The upper panel in Figure 3 displays the result along the $[0,0,\xi]$ direction
and similar data along the $[\xi,\xi,0]$ direction is shown in the bottom
panel.
It is clear that
spin waves are significantly damped at large wavevectors. Although still
relatively
well-defined  throughout the Brillouin zone in the $[0,0,\xi]$ direction
for both compounds,
the excitations are below the sensitivity of the measurements at
wavevectors beyond
(0.25,0.25,0) reciprocal
lattice units (rlu)
along the $[\xi,\xi,0]$ direction for NSMO.

To further investigate the wavevector dependence of the spin-wave
broadening and damping,
we plot in Figure 4
the intrinsic widths of the magnons along the $[0,0,\xi]$ direction.
The full width at half maximum (FHWM)
of the excitations $\Gamma$ shows a similar increase
at  wavevectors larger than $\xi\approx0.3$ rlu for all three manganites
\cite{note3}.
To determine whether such broadening is due to
the Stoner continuum, we note that at low temperatures,
the spin moment of itinerant electrons of ferromagnetic manganites is
completely
saturated and
the system is in the half-metallic state \cite{park}.
In this scenario of the DE model, there is complete separation of the
majority and
minority band by a large $J_H$. As a consequence, the Stoner continuum is
expected to lie
at an energy scale (2$J_H$) much higher than that of the spin-wave
excitations \cite{furukawa,okabe}.  For this reason, the observed
magnon broadening and damping
for lower $T_C$ manganites are unlikely to be due to Stoner continuum
excitations.

On the other hand, such behavior may be well understood if
one assumes a new spin-wave damping channel that is
related to a strong coupling between the conduction band
($e_g$) electrons and the cooperative oxygens in the Mn-O-Mn bond,
analogous to that of a dynamic
JT effect \cite{zhou}.
Although JT based electron-lattice coupling
is known to be important for the metal-to-insulator transition
at temperatures near and above $T_C$
\cite{millis,bishop}, such coupling may also be important to
understand the low temperature magnetic properties.

If electron-lattice coupling is indeed responsible for the observed
spin-wave broadening and damping, one would expect the presence of
such coupling in the lower $T_C$ samples that should be absent
in the higher $T_C$ materials.
Experimentally, there have been no reports of magnon-phonon coupling in
the higher $T_C$  $A_{0.7}B_{0.3}$MnO$_{3}$ \cite{martin,toby,endoh}.
For the DE ferromagnet
La$_{0.8}$Sr$_{0.2}$MnO$_3$ ($T_C=304$ K),
Moudden {\it et al.} \cite{moudden} have measured the spin-wave,
acoustic and optical phonon dispersions
from the zone center to the zone boundary along the $[0,0,\xi]$ direction.
 Phonon branches
were found to cross smoothly through the  magnon dispersion, suggesting
weak or no
magnon-phonon hybridization.   For the lower $T_C$ ferromagnet
LCMO, infrared reflectivity spectra show three zone center (${\bf q}=0$)
transverse optical (TO)
phonons located around 20.4 meV, 42 meV, and 73 meV \cite{kim}.
These three phonon modes were identified
as ``external'', ``bending'', and ``stretching'' modes
which correspond to the vibration of the La/Ca ions against the MnO$_6$
octahedron, the bending motion of the Mn-O-Mn bond,
and the internal motion of the
Mn ion against the oxygen octahedron, respectively.

To search for possible magnon-phonon
coupling in the lower $T_C$ manganites, we have
measured selected optical phonons in the LCMO single crystal.
 Figure 2 shows two longitudinal optical (LO)
phonon modes throughout the Brillouin zone as
solid symbols.
The metallic nature of the manganites at low
temperature requires the collapse of
the LO and TO splitting of the polar modes at ${\bf q}=0$. For LCMO,
this means that the two
LO modes observed by neutron scattering are likely to be the external and
bending modes identified by infrared reflectivity \cite{kim}.
At wavevectors larger than 0.3 rlu,
the dispersions of these two phonon modes
 are remarkably close to those of the magnons
along the $[0,0,\xi]$ and $[\xi,\xi,0]$
directions, suggesting that the softening of the spin-wave branches in Figure 2
is  due to magnon-phonon coupling.

In principle, the interaction between the magnetic moments and the lattice
can modify spin waves in two different ways.
First, the static lattice deformation induced
by the ordered magnetic moments may affect the anisotropy of the spin waves.
Second, the dynamic time-dependent modulations of
the magnetic moment may interfere with
the lattice vibrations, resulting in significant magneto-elastic
interactions or
magnon-phonon coupling.
One possible consequence of such coupling is to create energy
gaps in the magnon dispersion at the nominal intersections of the magnon
and phonon modes. However, our spin-wave dispersion data in Figure 2 show no
obvious evidence of any gap at the magnon and phonon crossing at
$\xi\approx 0.3$ rlu. Alternatively,
magnon-phonon coupling, present in all exchange coupled
magnetic compounds to some extent, may give raise to
spin-wave broadening \cite{lovesey}.
 In this scenario, the vibrations of the magnetic ions
about their equilibrium positions affect the exchange energy
through the spatial variation of the spin-spin exchange
coupling strength, which in turn leads to spin-wave broadening
at the magnon-phonon crossing points.
Generally, one would expect such coupling to be strong
for the lower $T_C$  $A_{0.7}B_{0.3}$MnO$_{3}$ manganites because of their
close
proximity to the charge-ordered insulating state \cite{hwang1}.
This is exactly what is observed for these materials at $\xi\ge 0.3$ rlu.
Constant-{\bf q} scans
in Figure 3 show significant broadening of the
spin waves from $\xi<0.3$ to $\xi\ge0.3$ rlu for LCMO and NSMO.
Similarly, Figure 4
reveals that
magnon widths increase considerably at wavevectors  $\xi \ge 0.3$ rlu
for all three manganites investigated, consistent with
the expectation of a strong magnon-phonon hybridization \cite{note2}.

We have discovered that
spin-wave softening and broadening
along the $[0,0,\xi]$ direction occur at the
nominal intersection of the magnon and optical phonon modes in lower $T_C$
$A_{0.7}B_{0.3}$MnO$_{3}$ manganites.
This result strongly suggests that magneto-elastic coupling is important
to the understanding of the low temperature
spin dynamics of $A_{0.7}B_{0.3}$MnO$_{3}$.
In the lower doping canted ferromagnet
La$_{0.85}$Sr$_{0.15}$MnO$_3$ \cite{hazuki},
much larger spin-wave broadening and damping
have been found at low temperatures \cite{deloc}.
Although the magnon dispersion relation in that system appears to be
consistent with the simple nearest-neighbor
Heisenberg Hamiltonian, the observation of strong
anisotropic spin-wave broadening is in
sharp contrast to the expectation of
the single-band DE
model where magnons in the ground state are eigenstates of the system
\cite{deloc}.
Therefore, it becomes clear that the
single-band DE mechanism cannot describe the
spin dynamics of La$_{0.85}$Sr$_{0.15}$MnO$_3$
and lower $T_C$ $A_{0.7}B_{0.3}$MnO$_{3}$ manganites,
not even in the low temperature ground
state. To understand the extraordinary
magnetic and transport properties of $A_{1-x}B_{x}$MnO$_{3}$, one must
explicitly consider
the close coupling between charge, spin, and lattice degrees of freedom
in these complex materials.

We thank H. Kawano, W. E. Plummer, S. E. Nagler, H. G. Smith, and X. D.
Wang for helpful discussions. This work was supported
by the US DOE under Contract No. DE-AC05-96OR22464  with Lockheed Martin
Energy Research Corporation and JRCAT of Japan.

\begin{figure}
\caption{Temperature dependence of the resistivity $\rho(T)$ for single
crystals used
in the neutron scattering experiments. The large drop in $\rho(T)$ corresponds
to the Curie Temperature ($T_C$) of our samples. They
are 301 K, 238 K, and 198 K for PSMO, LCMO, and NSMO, respectively. The
inset shows
the normalized resistivity $\rho(T)/\rho(0)$, it is clear that all three
samples
have the same $\rho(T)/\rho(0)$ temperature dependence below about 100 K.}
\label{autonum}
\end{figure}

\begin{figure}
\caption{Open symbols show magnon dispersions along the $[0,0,\xi]$,
$[\xi,\xi,0]$,
and $[\xi,\xi,\xi]$
directions for $\sim30\%$ manganites PSMO (open squares), LCMO
(open circles), and
NSMO (open down triangles) at 10 K.
The data for PSMO
are from Ref. [5]. The solid line is
a fit to a nearest-neighbor Hamiltonian assuming isotropic spin-waves for
$\xi<0.1$. Full symbols show selected LO phonon modes
collected along the reciprocal lattice directions as
specified in the legend. The rapid decrease of the manganese magnetic form
factor
at these large wavevectors
ensure that the scattering stem mostly from the lattice vibrations (phonons). }
\end{figure}

\begin{figure}
\caption{Constant-${\bf q}$ scans at selected wavevectors
for LCMO and NSMO at 10 K. The magnetic nature of the
signal was confirmed by measuring the temperature and wavevector dependence
of the scattering
(see, for example, Figure 2 of Ref. [5]).
The data were taken
with PG as monochromator and analyzer at a final neutron energy of $E_f
=13.6$ meV.
Analyzer turned background have been subtracted from the data. There is a
dispersionless
crystal electric field (CEF) level at $\sim$12 meV from Nd for NSMO.
The horizontal bars show the resolution along the scan direction. Solid lines
are Gaussians fits to the data.}
\end{figure}

\begin{figure}
\caption{The intrinsic magnon widths along the $[0,0,1+\xi]$ direction for
PSMO, LCMO, and
NSMO at 10 K. Significant broadening of the magnon widths is seen
at $\xi\approx 0.3$ rlu. The arrow indicates the crossing point of magnon
and phonon dispersions. Solid line is the guide to the eye.}
\end{figure}

\begin{references}
\bibitem{note1} Although the samples were orthorhombic and twinned, we
assume a cubic
lattice ($a=3.86$ \AA) for all three materials
because the spectrometer could not resolve the effects
of the twinning.
\bibitem{note3} To obtain $\Gamma$, we have
deconveluted the spin-wave widths from the instrumental resolution
calculated using the
spin-wave dispersion  at the measured wavevectors.
\bibitem{note2} In
principle, we should also observe a large spin-wave broadening at the
magnon-phonon
crossing points around 42 meV along the $[\xi,\xi,0]$ direction. However,
we were unable
to extract reliable magnon widths data  due to the weakness of the spin-wave
signal at these large energies (see Figure 3).
\end{references}
\end{document}